\documentclass[prl,letterpaper,twocolumn,nofootinbib,preprintnumbers,superscriptaddress]{revtex4-2} 

\pdfpageattr {/Group << /S /Transparency /I true /CS /DeviceRGB>>}

\usepackage[utf8]{inputenc}

\usepackage{titlesec}
\usepackage{dcolumn}
\usepackage{bm}
\usepackage{amsmath,amssymb,amsfonts}

\usepackage{colortbl}
\usepackage{xfrac}
\newcommand{\ra}[1]{\renewcommand{\arraystretch}{#1}}
\usepackage{booktabs}
\usepackage{graphicx}
\usepackage{color}
\usepackage{bbold}
\usepackage{lipsum}

\usepackage{slashed}
\usepackage[svgnames]{xcolor}
\usepackage[english]{babel}
\usepackage{blindtext}
\usepackage{microtype}
\usepackage{tikz}
\usepackage[most]{tcolorbox}
\usepackage{libertine}
\usepackage[colorlinks=true,linkcolor=blue,urlcolor=blue,citecolor=blue]{hyperref}
\usepackage{ifpdf}

\newtcbox{\mymath}[1][]{%
    nobeforeafter, math upper, tcbox raise base,
    enhanced, colframe=blue!30!black,
    colback=blue!30, boxrule=1pt,
    #1}

\usepackage{slashed}
\usepackage{selinput}
\SelectInputMappings{
    adieresis={ä},
    germandbls={ß}
}

\usepackage{libertine}

\newcommand{\Eq}[1]{Eq.~\eqref{#1}}

\newcommand{\Refs}[1]{Refs.~\cite{#1}}

\newcommand{\Fig}[1]{Fig.~{\ref{#1}}}

\newcommand{\Table}[1]{Table~\ref{#1}}

\renewcommand{\Ref}[1]{Ref.~\cite{#1}}

\newcommand{\vect}[1]{\boldsymbol{#1}}

\newcommand{\sh}[1]{\slashed{#1}}

\newcommand{\CDOT}{\mbox{$ \cdot $}}
\newcommand{\pslash}{\mbox{$\not \! p$}}
\newcommand{\aslash}{\mbox{$\not \! a$}}
\newcommand{\nslash}{\mbox{$\not \! n$}}

\newcommand{\Qperpslash}{\mbox{$\not \! \! Q_\perp$}}

\graphicspath{{.}{figures/}} 

\makeatletter
\newcommand*\if@single[3]{%
  \setbox0\hbox{${\mathaccent"0362{#1}}^H$}%
  \setbox2\hbox{${\mathaccent"0362{\kern0pt#1}}^H$}%
  \ifdim\ht0=\ht2 #3\else #2\fi
  }
\newcommand*\rel@kern[1]{\kern#1\dimexpr\macc@kerna}
\newcommand*\widebar[1]{\@ifnextchar^{{\wide@bar{#1}{0}}}{\wide@bar{#1}{1}}}
\newcommand*\wide@bar[2]{\if@single{#1}{\wide@bar@{#1}{#2}{1}}{\wide@bar@{#1}{#2}{2}}}
\newcommand*\wide@bar@[3]{%
  \begingroup
  \def\mathaccent##1##2{%
    \if#32 \let\macc@nucleus\first@char \fi
    \setbox\z@\hbox{$\macc@style{\macc@nucleus}_{}$}%
    \setbox\tw@\hbox{$\macc@style{\macc@nucleus}{}_{}$}%
    \dimen@\wd\tw@
    \advance\dimen@-\wd\z@
    \divide\dimen@ 3
    \@tempdima\wd\tw@
    \advance\@tempdima-\scriptspace
    \divide\@tempdima 10
    \advance\dimen@-\@tempdima
    \ifdim\dimen@>\z@ \dimen@0pt\fi
    \rel@kern{0.6}\kern-\dimen@
    \if#31
      \overline{\rel@kern{-0.6}\kern\dimen@\macc@nucleus\rel@kern{0.4}\kern\dimen@}%
      \advance\dimen@0.4\dimexpr\macc@kerna
      \let\final@kern#2%
      \ifdim\dimen@<\z@ \let\final@kern1\fi
      \if\final@kern1 \kern-\dimen@\fi
    \else
      \overline{\rel@kern{-0.6}\kern\dimen@#1}%
    \fi
  }%
  \macc@depth\@ne
  \let\math@bgroup\@empty \let\math@egroup\macc@set@skewchar
  \mathsurround\z@ \frozen@everymath{\mathgroup\macc@group\relax}%
  \macc@set@skewchar\relax
  \let\mathaccentV\macc@nested@a
  \if#31
    \macc@nested@a\relax111{#1}%
  \else
    \def\gobble@till@marker##1\endmarker{}%
    \futurelet\first@char\gobble@till@marker#1\endmarker
    \ifcat\noexpand\first@char A\else
      \def\first@char{}%
    \fi
    \macc@nested@a\relax111{\first@char}%
  \fi
  \endgroup
}
\makeatother

\begin{document}

\title{ Parton Decomposition of Nucleon Spin and Momentum: Gluons from Dressed Quarks}

\author{Peter~C.~Tandy}
\affiliation{Center for Nuclear Research, Department of Physics, Kent State University, Kent OH 44242 USA}

\affiliation{CSSM, Department of Physics, University of Adelaide, Adelaide SA 5005, Australia}

\begin{abstract}
The lowest two Mellin moments of hadronic Generalized Parton Distributions  are explored within a model that allows investigation of the inter-related quark and gluon contributions.  For light quarks their dynamical connection is strong due to quark dressing.  Our principal focus is the angular momentum $J$ of the nucleon.  This work employs and extends dynamical insights obtained from our recent model results for quark and gluon momentum fractions 
$\langle x \rangle_{\rm q/g}$ in both pion and nucleon.   The employed model  is based on the Rainbow-Ladder truncation of the Dyson-Schwinger equations of QCD.  
The special case of a 1-loop treatment of a single hadronic quark is used to motivate several insights and obtain initial estimates such as the Wilson line correction to the established Landau gauge model ($ -7\%$ for both $J_{\rm q}$ and $\langle x \rangle_{\rm q}$),  and the "binding gluon" contribution to  $\langle x \rangle_{\rm g}$ ($\leq 10\%$).  We obtain  the proton $J$ within 1\% after inclusion of the pion cloud mechanism to produce the sea.    The gluon second Mellin moments of the proton reflect similar dynamics;  ($\langle x \rangle_{\rm g}$, $J_{\rm g}$)  are  (26\% , 24\%)  at model scale, and  (40\%, 38\%) at $2$~GeV.  We also find  that $J_{\rm tot}$ is shared almost equally between the total orbital and total intrinsic spin contributions.  
\end{abstract}

\date{February 13, 2023}


\maketitle

\noindent\textbf{Introduction:} 
It is becoming increasingly evident that the gluon contribution to hadron properties and structure must be accounted for to complement the long-standing quark basis. 
Such a more complete picture is starting to emerge~\cite{Lin:2017snn,Lin:2020rut} from experimental and theoretical work on integral properties such as the quark and gluon parton contributions to the nucleon  angular momentum and lightcone momentum~\cite{Deka:2013zha,Yang:2016plb,Yang:2018bft,Fan:2022qve}.   The lattice-regulated approach to QCD calculations has in recent years moved beyond the low moments of quark parton distribution functions (PDFs) and is able to address the parton angular momentum contributions as well as a more complete momentum fraction $x$-dependence of PDFs, see e.g., \Refs{Sufian:2019bol,Gao:2020ito,Alexandrou:2017oeh,Alexandrou:2020sml}.
Understanding of how the total angular momentum of the nucleon is made from its constituent quarks and gluons has long been of high interest and a settled picture has not yet emerged.    The traditional modelling approach has been to begin with the quark spin-isospin SU(6) state and then to supplement it with selected relativistic mechanisms for orbital contributions without clear dynamical consistency.  The resulting nucleon $J$ was not gauge invariant and hence not an observable; also no guidance could be provided that way for the gluon contribution $J_{\rm g}$.

Here we investigate the  parton decomposition of nucleon $J$ directly from the leading-twist QCD unpolarized generalized parton distributions (GPDs) as formulated  by Ji~\cite{Ji:1996ek,Ji:1996nm}. 
The advantage in this approach is that the resulting  decompositions \mbox{$J = J_{\rm q} + J_{\rm g}$} and \mbox{$J_{\rm q} = L_{\rm q} +S_{\rm q}$} have established gauge invariant QCD matrix elements for the individual terms~\cite{Ji:1996ek}.   However separate QCD gauge-invariant matrix elements for each term of a gluon decomposition \mbox{$J_{\rm g} = L_{\rm g} + S_{\rm g}$}  have not been established~\cite{Ji:1996ek,Diehl:2003ny}.  Our main interest is the size of $J_{\rm q/g} $ at model scale, and the dominance of the gluon-in-quark effect from dressing. 

The GPDs are of broader interest~\cite{Diehl:2003ny} because they contain information on imagining of hadrons~\cite{Burkardt:2002hr}, their mass decomposition~\cite{Lorce:2018egm,Hatta:2018sqd}, and cross sections for hard exclusive reactions~\cite{Ji:1996nm,Radyushkin:1997ki} that can be measured at facilities such as Jefferson Lab and an Electron Ion Collider.  The lowest Mellin moments of hadron GPDs at $Q^2=0$ relate to static properties such as the electric charge, magnetic moment, axial charge, and quark spin $S_{\rm q}$.  More generally, the parton shares of total angular momentum of hadrons is  obtainable through the $Q$ dependence of the second Melin moments of GPDs. 
A Nambu-Jona-Lasinio model approach to GPDs has investigated the sharing of orbital angular momentum and intrinsic spin between the nucleon's valence u- and d-quarks~\cite{Freese:2020mcx}, however the gluon contribution at the intrinsic model scale was not accessible in that format.  Recently the light-front Hamiltonian diagonalization method has been extended to address the parton decomposition of nucleon spin while including an explicit one-gluon Fock space component~\cite{Xu:2022abw}.

We obtain the parton decomposition of the proton $J$ from the Dyson-Schwinger equation approach (DSE) to the second Mellin moments of quark and gluon GPDs.   In ladder-rainbow (RL) truncation  an infinite subset of gluon emission and absorption processes is included; the quark dressing mechanism, so strong because of dynamical chiral symmetry breaking,  is found to generate most of the gluon $J_{\rm g}$ and $\langle x \rangle_{\rm g}$ at the model scale. 
We use the same DSE-RL approach that has proven  to be very efficient for ground state masses, decay constants, and electromagnetic form factors~\cite{Bashir:2012fs,Cloet:2013jya,Tandy:2014hja,Horn:2016rip}.  It has been  especially accurate for light quark pseudoscalar and vector mesons~\cite{Maris:1999nt,Maris:2000sk} because their properties are strongly dictated by the dynamical breaking of chiral symmetry and vector current conservation, which is built into the approach.  It has been applied to pion, kaon and nucleon PDFs~\cite{Nguyen:2011jy,Chang:2014lva,Chang:2014gga,Chen:2016sno,Bednar:2018htv,Shi:2018mcb,Ding:2019lwe} mostly using the Ward Identity Ansatz to represent the relevant quark vertex for any PDF moment.  As discussed later, this Ward Identity vertex is accurate only for the lowest (quark number) moment, and does not distinguish the gluon-in-quark and quark-in-quark PDFs.   DSE approaches that do incorporate this distinction have been applied to $q(x)$ for the pion and nucleon~\cite{Bednar:2018mtf,Freese:2021zne}; the present work is an extension.  When needed, the quark and gluon helicities  can be obtained from the polarized PDF results 
available  within the present model~\cite{Freese:2021zne}.  

\smallskip 
\noindent\textbf{Parton Angular Momenta in the Nucleon:} 
Generalized parton distributions (GPDs) are defined via Poincar\'e-invariant matrix elements
of bilocal light cone correlators of fields.  The leading-twist, unpolarized quark GPDs of a nucleon are identified from the  frame-independent  matrix element~\cite{Diehl:2003ny}
\begin{align}
\mathcal{J}_{s', s}^{\rm q} = \int \frac{d\lambda}{4\,\pi}\,  e^{i x K\CDOT n \lambda} 
\left\langle P' \middle| \bar{q}(-\lambda \frac{n}{2})\,  \sh{n}\, W \,q(\lambda \frac{n}{2})
\middle| P \right\rangle_c.
\label{eq:JQFTq}
\end{align}
Here $P, s$ and $P', s'$ are the momentum and spin projection of the initial and final nucleon states ($s^\prime, s$ are to be understood on the right hand side above),  $W(\lambda, A)$ is the  Wilson line integral that restores  gauge invariance to the non-local matrix element, \mbox{$Q=P'-P$}, and    $K=\frac{P'+P}{2}$.  We employ throughout the  light-like longitudinal basis vector \mbox{$n = (1,\,\vect{0}_T,\,-1)$}, and note that \Eq{eq:JQFTq} takes the same form if a factor (e.g., $1/\sqrt{2}$)  is absorbed into $n$.   Besides momentum fraction $x$ and $Q$, the matrix elements $\mathcal{J}_{s' s}^{\rm q}(x,\xi,Q)$ also depend upon \mbox{$\xi = - Q \CDOT n/(2K \CDOT n)$}.   The corresponding gluon matrix elements, in the Ji convention~\cite{Ji:1996ek,Diehl:2003ny},  are  
\begin{align}
\mathcal{J}_{s', s}^{\rm g} = \int \frac{d\lambda}{4\,\pi}  \frac{e^{i x K \CDOT n \lambda} }{x K\CDOT n} 
\left\langle P' \middle| G^{n\, \alpha}(-\lambda  \frac{n}{2})\,W\,G_{\alpha \,n}(\lambda  \frac{n}{2})
\middle| P \right\rangle_c,
\label{eq:JQFTg}
\end{align}
where we use the notation \mbox{$G^{n\, \alpha}(z) = n_\nu \,G^{\nu\, \alpha}(z)$}. The nucleon GPDs  $H^{\rm q/g}(x,\xi,Q^2)$ and $E^{\rm q/g}(x,\xi,Q^2)$,  can be  identified from the general form \begin{align}
\mathcal{J}_{s', s}^{\rm q/g}  =  \frac{M}{K \CDOT n} \,\bar{u}_{s'}(\!P'\!)  \left[  \sh{n}\, H^{\rm q/g}
  +   \frac{i\sigma^{n \alpha}Q_\alpha}{2M}\,  E^{\rm q/g}   \right] u_{s}(\!P\!) .
\label{eq:gen_form_JQFT}
\end{align}
Here, for convenience,  we have introduced spinors normalized as \mbox{$\bar{u}_{s^\prime} u_s = \delta_{s^\prime s} $}. 

The separate quark and gluon contributions to $J^z$ (hereafter denoted $J$) are given by
\begin{align}
J_{\rm q/g} = \frac{1}{2} \int dx \, x\, \Big( H^{\rm q/g}(x,0,0) +  E^{\rm q/g}(x,0,0)  \Big) \, ,
\label{eq:ji}
\end{align}
where  the integral terms e.g., \mbox{$A^{\rm q/g}(0) =  $}  \mbox{$\int dx \, x\, H^{\rm q/g}(x,0,0)$}, are the parton components of the gravitational form factor strengths. The first term also yields the momentum fractions  \mbox{$ \langle x \rangle_{\rm q/g} = A^{\rm q/g}(0) $}, 
which are available from prior work within the present approach~\cite{Freese:2021zne}.   To produce $E^{\rm q/g}(x)$, and thus  $J_{\rm q/g}$, a more general approach involving the $Q$ dependence is needed. 

To  directly extract $J_{\rm q/g}$ via \Eq{eq:ji}, it is convenient to adopt the 
special frame where \mbox{$P = K - Q/2$}, \mbox{$P' = K + Q/2 $},  with $Q$ purely transverse and \mbox{$K = (K^0, \vec{0})$}.  Thus \mbox{$\xi = 0= n \CDOT Q = K \CDOT Q$}, and the mass-shell condition for $P, P^\prime$  yields \mbox{$M/K \CDOT n = (1+Q_{\perp}^2/(4 M^2) )^{-1/2}$}.  (For \mbox{$ Q_{\perp} \to 0 $} this special frame is the target rest frame.) The second Mellin moment of the spin-flip matrix element, after retention of just the leading linear dependence upon $Q_\perp$, yields 
\begin{align}
\int dx \, x\, \mathcal{J}_{\uparrow \downarrow}^{\rm q/g}(x,0,Q) =  \left[A_{\rm q/g}(Q^2) + B_{\rm q/g}(Q^2) \right]\, \frac{ i\, Q_\perp}{2 M} \,,
\label{eq:AplB_id}
\end{align}
thus allowing the direct evaluation 
\begin{align}
J_{\rm q/g} = \lim_{Q \to 0} \, N_{\rm J}(Q_\perp)  \int dx \, x\, \mathcal{J}_{\uparrow \downarrow}^{\rm q/g}(x,0,Q)  \,,
\label{eq:J_id}
\end{align}
where \mbox{$N_{\rm J}(Q_\perp) = -i M/Q_\perp$}.  Note that only the linear dependence upon small $ Q_\perp$ need be retained and thus $M/K \CDOT n$ in \Eq{eq:gen_form_JQFT} makes no contribution to $J_{\rm q/g} $.  

\smallskip
\noindent\textbf{Dynamical Approach:} 
The previous definitions employed Minkowski metric; from here on we adopt Euclidean metric in order to apply the Rainbow-Ladder truncation of the DSE approach to non-perturbative aspects of hadron physics.\footnote{In Euclidean metric we employ $a_4 = i a^0$ for any space-time vector, including $n$, while  \mbox{$\{\gamma_\alpha, \, \gamma_\beta\} =2 \delta_{\alpha \beta} $}  with $\gamma_4 = \gamma^0$.   Hence \mbox{$\aslash \to -i \aslash $} while \mbox{$a \CDOT b \to - a \CDOT b $}.  }  This has successfully described many hadron properties~\cite{Bashir:2012fs,Cloet:2013jya,Tandy:2014hja,Horn:2016rip}, in particular it was previously used to explore dynamical relations between quark and gluon unpolarized PDFs of the proton~\cite{Freese:2021zne}.  The DSE-RL truncation of the second Mellin moment of the spin-flip QCD matrix element in \Eq{eq:JQFTq} for the quark, and in \Eq{eq:JQFTg} for the gluon,  yields  
\begin{align}
\mathcal{J}_{\uparrow, \downarrow}^{\rm q/g} = {\rm tr}\! \int_p^R \!\!\! S(p_+\!)\, \Gamma_{\rm q/g}^{(1)}(p,Q_\perp\!)\, S(p_-\!) {\mathcal M}_{\uparrow, \downarrow }(p,Q_\perp;\!P), 
\label{eq:DSE_spinF_N_q}
\end{align}
where \mbox{$p_\pm = p \pm Q_{\perp}/2$}, the trace is over Dirac and color indices, and $\int_p^R$ represents $\int \frac{d^4p}{(2 \pi)^4}$ with smooth regularization.  The nucleon spin-flip matrix element ${\mathcal M}_{\uparrow, \downarrow } $ is described later.

In \Eq{eq:DSE_spinF_N_q} $ \Gamma_{\rm q/g}^{(1)}(p,Q_\perp)$ are DSE-RL dressed vertices that carry the  second Mellin moments generated from the quark-in-quark and gluon-in-quark aspects.   The strength of dynamical chiral symmetry breaking for light quarks suggests this sourcing of the gluon parton behavior from the dressing mechanism; the  results justify this.   The defining Bethe-Salpeter equation is~\cite{Nguyen:2011jy,Bednar:2018mtf,Freese:2021zne}
\begin{align}
&\Gamma_{\rm q/g}^{(m)}(p,Q_\perp) =  \Gamma^{(m)}_{\rm q/g,D} (p,Q_\perp) \nonumber \\
& - \int_\ell^R \frac{\lambda^a}{2} \gamma_\mu \mathcal{K}_{\mu \nu}(q)\, S(\ell_+)\, \Gamma_{\rm q/g}^{(m)}(\ell,Q_\perp)\, S(\ell_-)\, \frac{\lambda^a}{2} \gamma_\nu~, 
\label{eq:Gamma_BSE}
\end{align}
where \mbox{$q = p - \ell $}, and  $\mathcal{K}_{\mu \nu}(q)$ is the DSE-RL kernel given later.  The inhomogeneous term for the quark-in-quark vertex  is \mbox{$\Gamma^{(m)}_{\rm q,D} (p,Q_\perp) =  $}  \mbox{$Z_2\,(-i \nslash)\,(\frac{ p \CDOT n}{K \CDOT n})^m$}.  This arises as follows:  the QCD matrix element in \Eq{eq:JQFTq} has longitudinal quark momentum fixed by $x$, conversion to a full momentum  integral produces \mbox{$\int d^4k \, \delta(x - k \CDOT n/K \CDOT n)\, F(k)$}, then integration over $x$  gives a factor $(k \CDOT n/K \CDOT n)^m$ occurring in any Feynman integral expression for $\langle x^m \rangle$, see \Refs{Nguyen:2011jy,Bednar:2018mtf,Freese:2021zne}.

The inhomogeneous term for the gluon-in-quark  vertex  is 
\begin{align}
\Gamma^{(m)}_{\rm g,D} (p,Q_\perp\!) \!= \!\frac{4}{3} \!\int_{\rm q}^R \!\!\big(\!\frac{q \CDOT n}{K \CDOT n}\!\big)^m  \gamma_\mu \, \hat{{\mathcal D}}_{\mu \nu}(q,Q_\perp\!) \,S(p\!-\!q) \, \gamma_\nu ,
\label{eq:Inhom_GVertm}
\end{align}
where $S(p\!-\!q)$ is the quark propagator and \mbox{$ \hat{{\mathcal D}}_{\mu \nu}(q,Q_\perp\!) = $}  \mbox{${\mathcal K}_{\rm g}(q^2)\, \hat{D}_{\mu \nu}(q,Q_\perp\!)  $} with the kinematic tensor  $\hat{D}_{\mu \nu}(q,Q_\perp\!)$ given in an Appendix.  The gluon-in-quark PDF kernel ${\mathcal K}_{\rm g}(q^2)$ is a non-perturbative generalization of the bare \mbox{${\mathcal K}^0_{\rm g}(q^2) = g^2/q^4$} and is specified later.  For application to $J_{\rm g}$ only the final linear  $Q_{\perp}$ dependence in \Eq{eq:Inhom_GVertm} is needed and higher order contributions have been discarded.

\smallskip
\noindent\textbf{Interaction Kernels:} 
For realistic numerical work, the dressed quark propagator $S(p)$ is obtained from the quark Dyson-Schwinger equation of QCD in Rainbow-Ladder truncation, which is 
\begin{align}
S^{-1}(p) = Z_2\,S_0^{-1}\!(p) - 
\!\int_k^R \!\frac{\lambda^a}{2} \gamma_\mu \,\mathcal{K}_{\mu \nu}(q)\,  S(k)\, \frac{\lambda^a}{2} \gamma_\nu,
\label{SigmaRL}
\end{align}
where \mbox{$ S_0^{-1}(p) = i \slashed{p} + Z_m m_r $}.  The general form of the solution is 
$S^{-1}(p)  = i \pslash \,A(p^2, \mu^2) + B (p^2, \mu^2)$, and the renormalization constants ($Z_2$, $Z_m$) produce \mbox{$A \to 1$} and \mbox{$B \to m_r$} at the renormalization scale $\mu$.  The  standard Landau gauge DSE-RL interaction kernel~\cite{Maris:1999nt,Nguyen:2011jy,Qin:2011dd} that generates quark propagators, BSE vertices and meson bound states is \mbox{$ \mathcal{K}_{\mu \nu}(q) = {\mathcal K}(q^2) \,T_{\mu \nu} (q)$} with \mbox{$T_{\mu \nu}(q) = \delta_{\mu \nu} - q_\mu q_\nu /q^2 $} and ${\mathcal K}(q^2)$ is the model's non-perturbative generalization of the bare coupling \mbox{${\mathcal K}^0(q^2) = g^2/q^2$}.  The model kernel is 
\begin{align}
{\mathcal K}(q^2) = D_{\rm RL}^2  \, {\rm e}^{-q^2/\omega^2} 
+ {\mathcal F}(q^2)\, 4\pi\, \tilde{\alpha}_s(q^2)  \, .
\label{eq:RL_Kernel}
\end{align}
Here $\tilde{\alpha}_s(q^2)$ denotes a continuation of the 1-loop $\alpha_s(q^2)$ to provide smooth non-singular coverage for the entire domain of $q^2$.   The first term of \Eq{eq:RL_Kernel} implements the infrared enhancement due to dressing effects, while the second term, with \mbox{${\cal F}(q^2)=$} \mbox{$(1 - \exp( -q^2/(1~{\rm GeV^2}) ) )/q^2 $}, connects smoothly with the 1-loop renormalization group behavior of QCD.  The DSE-RL kernel correlates a large amount of hadron physics~\cite{Bashir:2012fs,Cloet:2013jya,Tandy:2014hja,Horn:2016rip}.   Details and parameters are given in an Appendix.

The gluon-in-quark PDF kernel ${\mathcal K}_{\rm g}(q^2)$ that drives the corresponding quark vertex via \Eq{eq:Inhom_GVertm} has a 1-loop renormalized UV behavior related to that of the BSE kernel ${\mathcal K}(q^2)$ which is linked by global symmetries to the dressed quark Dyson-Schwinger equation.     Multiplicative renormalizability links the large  renormalization scale dependence of propagators and vertex functions to their ultraviolet momentum dependence, for example see \Ref{Maris:1997tm}.  The BSE-RL kernel for the interaction of 2 quark currents collects the ultraviolet 1-loop momentum dependence from \mbox{$Z_2^2\, Z_3/Z_{1F}^2 \to $}  \mbox{$4 \pi\,\alpha_s(q^2) \to 4 \pi^2 \, \gamma_m /{\rm ln}(q^2/\Lambda_{\rm QCD}^2)$},  where \mbox{$ \gamma_m = 4/\beta_0 = 12/(33-2\, N_f)$}.   For ${\mathcal K}_{\rm g}(q^2)$ there is an extra dressed gluon propagator; the deep ultraviolet behavior is characterized by an extra factor $Z_3(q^2,\Lambda^2)/q^2$. 

We take ${\mathcal K}_{\rm g}(q^2)$ in the form~\cite{Freese:2021zne}
\begin{align}
{\mathcal K}_{\rm g}(q^2) = D_{\rm gg}^4  \, {\rm e}^{-q^2/\omega_{\rm g}^2}
+ {\mathcal F}^2(q^2)\, 4\pi\, \tilde{\alpha}_s(q^2) \,\tilde{Z}_3(q^2) \, ,
\label{eq:CutG_Kernel}
\end{align}
where  $\tilde{Z}_3(q^2)$ denotes a continuation of the corresponding 1-loop $q^2$ dependence after the regularization mass scale $\Lambda$ has been absorbed into the definition of $\Lambda_{\rm QCD}$.    Details and parameters are given in an Appendix.

For the second Mellin moments of the GPDs in Eqs.\, (\ref{eq:JQFTq},\ref{eq:JQFTg}) in an arbitrary gauge, the Wilson line integral is unity for gluons but not so for quarks.  This difference is due to the extra factor of $1/x$ in the gluon matrix element.  Thus $\langle x \rangle_{\rm g}$ and $J_{\rm g}$ are  gauge invariant, but the quark contributions are not.   The DSE-RL approach to hadrons is established in Landau gauge and the Wilson line integral has usually been ignored because it is not easily accommodated  into integral equations that sum Feynman diagrams.  
\begin{figure}[tbp]
\vspace*{-34mm}\centering\includegraphics[width=0.9\columnwidth,height=120mm]{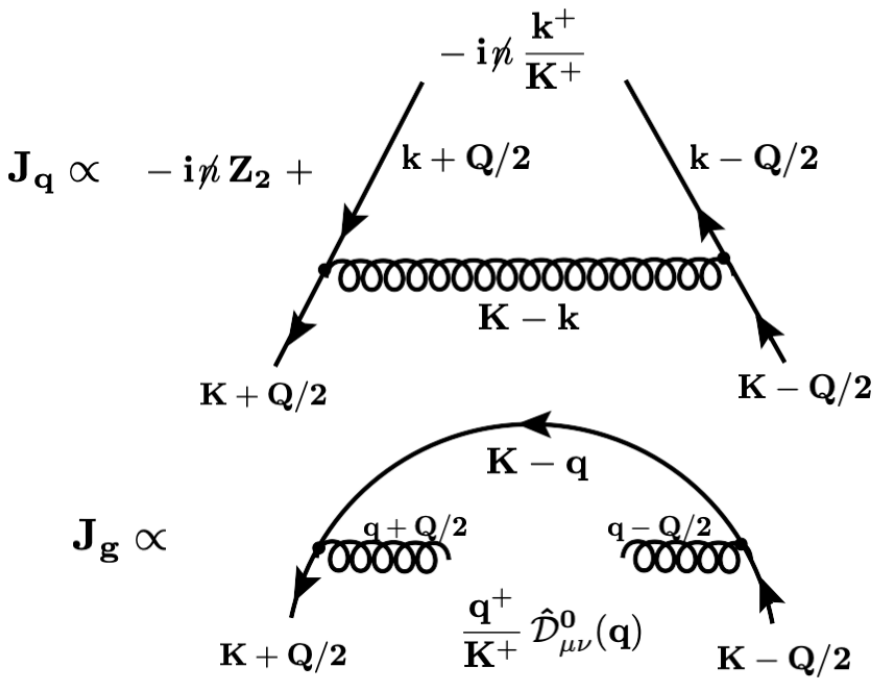} 
\vspace*{-38mm}\caption{Illustrations of the expressions used to obtain $J_{\rm q}$ and $J_{\rm g}$ for a quark target treated at 1-loop in QCD.   $J_{\rm q}$ in the top part, and $J_{\rm g}$ in the bottom part,  employ vertices given in Eqs.\, (\ref{eq:Gam_q_1loop_LC},\ref{eq:Gam_g_1loop} ) as per \protect\Eq{eq:Jq/g_1loop} and the discussion immediately following it. 
} 
\label{fig:1-loop_J_Diags}  
\end{figure}

\smallskip 
\noindent\textbf{Simple Case Insight} \mbox{\underline{ {\it  i) A Quark Dressed at 1-loop}:} }
The $\langle x \rangle_{\rm q/g}$ for a quark treated at 1-loop as a hadron with momentum $P$, follow from the QCD matrix elements in Eqs.\,(\ref{eq:JQFTq}, \ref{eq:JQFTg}) after the limit \mbox{$Q \to 0~, K \to P$}  in the quark target rest frame. The associated quark-in-quark and gluon-in-quark mechanisms are illustrated in \Fig{fig:1-loop_J_Diags}.   At 1-loop $\langle x \rangle_{\rm g}$ comes only from the dressing mechanism,  the Wilson line makes no contribution to it,  and the gauge-invariant result is \mbox{$\langle x \rangle_{\rm g} = $} \mbox{$ \bar{u}_{1} \,  \Gamma^{(1)}_{\rm g}(P) \, u_{1}$}.  (We continue  with spinor normalization \mbox{$\bar{u}_1(P) \,u_1(P)  = 1$}.)   
The relevant vertex here is the bare coupling version of that defined in \Ref{Freese:2021zne}, namely 
\begin{align}
\Gamma^{(1)}_{\rm g}(P)=   \frac{4}{3}  \int_{\rm q}^R \big( \frac{q \CDOT n}{P \CDOT n } \big)\, \gamma_\mu \, \hat{{\mathcal D}}^0_{\mu \nu}(q) \,S_0(P-q) \, \gamma_\nu~,
\label{eq:Gam_g_1loop}
\end{align}
where   $q \CDOT n/P \CDOT n$ is the gluon-in-quark momentum fraction,  $S_0$ is the bare quark propagator, and
the bare kernel \mbox{$\hat{{\mathcal D}}^0_{\mu \nu}(q) = $} \mbox{$ {\mathcal K}^0_{\rm g}(q^2)\, \hat{D}_{\mu \nu}(q) $} is the \mbox{$ Q_{\perp} \to 0 $}  limit of the quantity $\hat{{\mathcal D}}_{\mu \nu}(q,Q_{\perp})$ introduced in \Eq{eq:Inhom_GVertm} and needed to obtain $J_{\rm q}$.  In particular the tensor here,
\begin{align}
\hat{D}_{\mu \nu}(q) =  2\, q \CDOT n \,\Big( \delta_{\mu \nu} - \frac{ n_\mu  \,q_\nu + q_\mu \, n_\nu }{q \CDOT n}  
+ \frac{n_\mu \, q^2 \, n_\nu }{(q \CDOT n)^2} \Big)~,
\label{eq:D_explicit}
\end{align}
is $\hat{D}_{\mu \nu}(q,Q_{\perp} \to 0)$ given in an  Appendix.  

For $\langle x \rangle_{\rm q}$ we firstly choose light-cone gauge (in which the Wilson line is unity) and find \mbox{$\langle x \rangle_{\rm q} = $}  \mbox{$\bar{u}_{1}  \Gamma^{(1)}_{\rm q}(P) \, u_{1}$} with vertex
\begin{align}
\Gamma&^{(1)}_{\rm q}(P)  =  Z_2\,(-i \nslash )  \nonumber \\
& - \frac{4}{3}  \int_k^R \big( \frac{k \CDOT n}{P \CDOT n } \big)\, \gamma_\mu \, \hat{ {\mathcal K} }^0_{\mu \nu}(P-k) \;( n \CDOT \partial_k S_0(k)) \; \gamma_\nu ~, 
\label{eq:Gam_q_1loop_LC}
\end{align}
where the bare kernel  \mbox{ $\hat{ {\mathcal K} }^0_{\mu \nu}(q) = $} \mbox{$ {\mathcal K}^0(q^2) \, \hat{T}_{\mu \nu}(q)$ } employs \mbox{$\hat{T}_{\mu \nu}(q) = \delta_{\mu \nu}\! -\! (q_\mu n_\nu \!+\! n_\mu q_\nu)/q \CDOT n $}.  
The sum of these vertices produces
\begin{align}
\Gamma^{(1)}_{\rm q}(P)  + \Gamma^{(1)}_{\rm g}(P) =   - n \CDOT \partial_P\, S_1^{-1}(P) = \Gamma_{\rm 1,W}(P) ~, 
\label{eq:Gam_q_g_Sum_1loop_LC}
\end{align}
where the 1-loop propagator $S_1(P)$ has absorbed the renormalization constant $Z_2$, and $ \Gamma_{\rm 1,W}(P)$ is the Ward Identity vertex that implements vector current conservation at this level.  This vertex sum rule can be verified through integration by parts after noting  \mbox{$ - n \CDOT \partial_{\rm q} \, \hat{ {\mathcal K} }^0_{\mu \nu}(q) =$} \mbox{$ \hat{{\mathcal D}}^0_{\mu \nu}(q)  $}.        
The  Ward identity vertex aggregates the quark-in-quark and the gluon-in-quark vertices and produces the sum rule \mbox{$\langle x \rangle_{\rm q} +  \langle x \rangle_{\rm g} = 1$} at 1-loop.  
If the  quark vertex were changed to Landau gauge, the resulting $\langle x \rangle_{\rm q}$ is larger than the correct result $1 - \langle x \rangle_{\rm g}$ and the difference measures the Wilson line effect.

\begin{table}[h]
\ra{0.9}

\begin{tabular}{c|c|ccc}\hline
                  & $\mu~(GeV)$  & \hspace{2mm} $2\,\langle x \rangle_{\rm u_v} $ \hspace{1mm}   &  \hspace{1mm} $ \langle x \rangle_{\rm sea} $  & \hspace{2mm} $\langle x \rangle_{\rm g} $ \hspace{2mm}    \\
\hline
\rule{0em}{3ex}    
    $\pi$: \hspace*{4mm} Here  \hspace*{4mm}       & 0.9    & 0.602    & 0.141  & 0.257            \\   
\hline
\rule{0em}{3ex}    
    $\pi$: \hspace*{4mm} Here  \hspace*{4mm}     & $\sqrt{10}$    & 0.453                  &  0.152                    & 0.395    \\        
    \hspace*{5mm}JAM~\cite{Barry:2018ort}       & $\sqrt{10}$   & $0.45 \pm 0.01$  &  $0.17 \pm 0.01$   & $0.37 \pm 0.02$        \\                  
\hline 
\end{tabular}
\begin{tabular}{c|c|cccc}
                  & $\mu~(GeV)$  & \hspace{1mm} $\langle x \rangle_{\rm u_v} $ \hspace{1mm}  & \hspace{1mm} $\langle x \rangle_{\rm d_v} $ \hspace{1mm} &  \hspace{1mm} $ \langle x \rangle_{\rm sea} $  &   \hspace{1mm}$\langle x \rangle_{\rm g} $ \hspace{1mm}    \\
\hline
\rule{0em}{3ex}    
    N: \hspace*{3mm} Here  \hspace*{4mm}       & 0.64    & 0.413  & 0.175       &  0.160  &  0.252          \\
\hline
\rule{0em}{3ex}                                         
    N: \hspace*{3mm} Here  \hspace*{4mm}       & $\sqrt{5}$   & 0.265  & 0.112     &  0.198  &  0.425     \\
 NNPDF3.0,~\Ref{Ball:2014uwa}    & $\sqrt{5}$   & 0.273  &  0.111      &  0.175  &  0.441            \\
\hline
\end{tabular}
\caption{ Landau gauge DSE-RL model results  for $\langle x \rangle$ that now account for the estimated Wilson line effects at the revised model scales $\mu_0$.  After NLO scale evolution, values at higher scales are compared to global data analysis results.   }
\label{Pion-N_m1_mu0_mu}
\end{table}

The 1-loop truncation of  Eqs.\,(\ref{eq:JQFTq}, \ref{eq:J_id}) for a dressed quark at 1-loop yields the quark-in-quark and gluon-in-quark $J$ contributions 
\begin{align}
J_{\rm q/g} =  \lim_{Q \to 0} \, N_{\rm J}(Q_\perp) \left[ \bar{u}_{\uparrow}(P') \, \Gamma^{(1)}_{\rm q/g}(P^\prime,P)\,  u_{\downarrow}(P) \right] ~,
\label{eq:Jq/g_1loop}
\end{align}
as illustrated in  \Fig{fig:1-loop_J_Diags}.
Here $J_{\rm g}$ is gauge invariant,  and $\Gamma^{(1)}_{\rm g}(P^\prime,P)$ is given by \Eq{eq:Gam_g_1loop} with 
$\hat{{\mathcal D}}^0_{\mu \nu}(q) $ replaced by $\hat{{\mathcal D}}^0_{\mu \nu}(q,Q_{\perp})$ given in an Appendix.   For $J_{\rm q}$, the associated quark vertex   $\Gamma^{(1)}_{\rm q}(P^\prime, P) $ is the same as in \Eq{eq:Gam_q_1loop_LC} except with the replacement \mbox{$ n \CDOT \partial_k S_0(k) \to S_0(k_+)\,(-i \nslash ) \, S_0(k_-) $}, where \mbox{$k_\pm = k \pm Q_\perp/2 $}. 
Only the $Q_\perp$ dependence  of $ u_{\downarrow}(P)\, \bar{u}_{\uparrow}(P') $ makes a contribution to the Dirac trace in  \Eq{eq:Jq/g_1loop},  and  we set \mbox{$Q_{\perp} \to 0 $} everywhere else.   Thus we can set \mbox{$\Gamma^{(1)}_{\rm q/g}(P^\prime, P) \to \Gamma^{(1)}_{\rm q/g}(P) $} given in Eqs.\,(\ref{eq:Gam_g_1loop}, \ref{eq:Gam_q_1loop_LC}). 
With the light-cone gauge choice for $J_{\rm q}$ we again have \Eq{eq:Gam_q_g_Sum_1loop_LC} which produces 
\mbox{$J_{\rm q} +  J_{\rm g} = 1/2$} for a dressed quark at 1-loop. 

\smallskip 
\noindent\textbf{Simple Case Insight} \mbox{\underline{ {\it  ii) Estimated Wilson Line Effect}:}} 
For a single quark at 1-loop, the differences between the correct quark results ($1 -  \langle x \rangle_{\rm g}$, $1/2 -  J_{\rm g}$)  and the corresponding Landau gauge values ($\langle x \rangle_{\rm q}$, $J_{\rm q}$) are measures of the Wilson line corrections.  Effects of realistic running coupling, such as occur in the later DSE-RL application, significantly influence the corrections.  
For this dynamical 1-loop estimate we employ \mbox{$ {\mathcal K}^0(q^2) \to {\mathcal {\bar K}}(q^2)$} in \Eq{eq:Gam_q_1loop_LC}, and \mbox{$ {\mathcal K}_{\rm g}(q^2) \to {\mathcal {\bar K}}_{\rm g}(q^2)$} in \Eq{eq:Gam_g_1loop} and vary the strength parameters  corresponding to  
Eqs.\,(\ref{eq:RL_Kernel} ,\ref{eq:CutG_Kernel})  to simulate the DSE-RL model results at model scale \mbox{$\mu_0 = 0.64$}~GeV. 
Values are found to be \mbox{$\bar D_{\rm RL} =25.04$} and \mbox{$\bar D_{\rm gg} =3.52$} while approximating the quark propagator amplitudes  via \mbox{$A \sim 1.4 $} and \mbox{$B/A \sim 0.4 $}.  The integrals are regularized in the infrared and ultraviolet via the proper time method~\cite{Ebert:1996vx} with \mbox{$\Lambda_{\rm IR} = 0.05~{\rm GeV}$} and \mbox{$\Lambda_{\rm UV} = 20~{\rm GeV}$}.  The resulting second Mellin moments in both light-cone and Landau gauges are quite stable and are summarized by 
\begin{align}
\begin{tabular}{cc|cc|cc|cc} 
               $\langle x \rangle_{\rm g}$  &  \hspace{0mm} $ \langle x \rangle^{\rm LC}_{\rm q} $ \hspace{0mm}   &  \hspace{0mm} $\langle x \rangle^{\rm LG}_{\rm q}$   \hspace{0mm}   &    $ \langle x \rangle^{\rm W}_{\rm q} $\hspace{1mm}      &  \hspace{3mm} $J_{\rm g}$   \hspace{1mm}   &    $ J^{\rm LC}_{\rm q}  $\hspace{1mm}        &  \hspace{1mm}   $ J^{\rm LG}_{\rm q} $  \hspace{1mm}        &  \hspace{1mm}   $ J^{\rm W}_{\rm q} $           \\
\hline
                    0.233     &     0.767   &   0.821    &    $-6.5\%$    &  0.120       &   0.380   &  0.411    &  $-7.5\%$  \\
\end{tabular}~.
\label{eq:1loop_Wilson_estm}
\end{align}
The estimated  Wilson line correction to the Landau gauge quark results for second Mellin moments  is typically -7\%.

\smallskip 
\noindent\textbf{Simple Case Insight} \mbox{\underline{ {\it  iii) "Binding Gluon" Part of $\langle x \rangle_{\rm g}$}:} }
If the 1-loop vertices in Eqs.\, (\ref{eq:Gam_g_1loop},\ref{eq:Gam_q_1loop_LC}), and the dynamical relation between them, are generalized to ladder-rainbow summations, the vertex sum  \Eq{eq:Gam_q_g_Sum_1loop_LC} generalizes to  
\begin{align}
\Gamma^{(1)}_{\rm q}(p) + \Gamma^{(1)}_{\rm g}(p) = \big(\frac{p \CDOT n}{P \CDOT n}\big)\, \Gamma_{\rm W}(p)~,
\label{eq:q+g_m1_Vertex}
\end{align}
for a dressed valence quark of momentum $p$ within a hadron of momentum $P$.  Here \mbox{$\Gamma_{\rm W}(p) =  -n \CDOT \partial_p\, S^{-1}(p)$} is the DSE-RL Ward Identity vertex.  The use of \Eq{eq:q+g_m1_Vertex} in the hadron matrix element, summed over the valence quarks, aggregates the quark-in-quark and gluon-in-quark contributions to produce \mbox{$\langle x \rangle_{\rm Ward} = $}  \mbox{$ \langle x \rangle_{\rm q} + \langle x \rangle_{\rm g_D} $}, where we use $g_D$ to denote its gluons from quark dressing.  For example, a $\pi^+$ at DSE-RL level yields  
\begin{align}
\langle x \rangle_{\rm Ward}  =   {\rm tr} & \int_p^R  \big(\frac{p \CDOT n}{P \CDOT n}\big)\,  \bar{\Gamma}_\pi 
\, {S_{\rm u}}^\prime(p)  \,\Gamma_\pi\, S_{\rm d}(p-P) \nonumber \\
& +   \big(\frac{(p-P) \CDOT n}{P \CDOT n}\big)\, \bar{\Gamma}_\pi \, S_{\rm u}(p)\,\Gamma_\pi\, {S_{\rm d}}^\prime(p-P) ~,
\label{eq:pi_sum<x>}
\end{align}
\begin{figure}[tbh]
\vspace*{-39mm}\hspace*{1mm}\centering\includegraphics[width=\columnwidth,height=108mm]{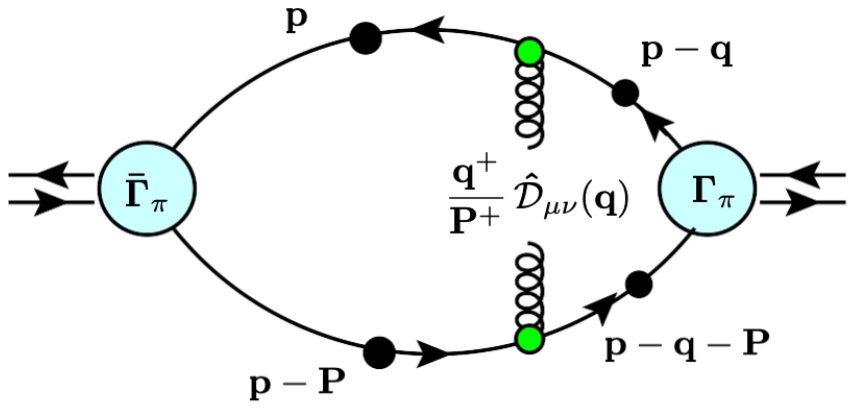}
\vspace*{-43mm}
\caption{ Diagrammatic illustration of the "binding" gluon contribution to the pion $\langle x \rangle_{g_{B}}$ as expressed by \Eq{eq:<x>gB_2loop}.
}
\label{fig:<x>gBind_Pi_m1}  
\end{figure}
where $ \Gamma_\pi$ is properly normalized and \mbox{$S^\prime(p) =  n \CDOT \partial_p\, S(p)$}.   In general the Bethe-Salpeter vertices  $\bar{\Gamma}_\pi $, $\Gamma_\pi $ have arguments $(q,-P)$ and $(q,P)$ respectively, where \mbox{$q = p-P/2 $} is the $\bar q q$ relative momentum.  Firstly consider a point pion limit where each $\Gamma_\pi $ is independent of $q$ (e.g., a Nambu-Jona-Lasinio model).  Then integration by parts applied to ${S_{\rm u}}^\prime(p) $ produces $ - {S_{\rm d}}^\prime(p-P) $ and cancels out the $p \CDOT n/P \CDOT n$ part of  the second term; the remainder is unity via the normalization condition for $\Gamma_\pi$~\cite{Maris:1999nt}, which is equivalent to \mbox{$\langle x^0 \rangle_{\rm u} = 1$}.   The point limit produces \mbox{$\langle x \rangle_{\rm Ward} = 1 $}.  However the $q$-dependence of a finite-size $\Gamma_\pi(q,P) $ generates an extra term, and the general result is  \mbox{$\langle x \rangle_{\rm Ward} = 1 - \langle x \rangle_{\rm g_B} $} where 
\begin{align}
\langle x \rangle_{\rm g_B} =   {\rm tr}  \int_p^R 
 \big(\frac{p \CDOT n}{P \CDOT n}\big)\, n \CDOT \partial_p\, \bar{\Gamma}_\pi \, S_{\rm u}(p^\prime) \, \Gamma_\pi \,  
 S_{\rm d}(p^\prime \!-\!P)\,|_{p^\prime \to p}~.
\label{eq:<x>gB_1loop_int}
\end{align}
When the ladder Bethe-Salpeter equation is used to resolve each $ n \CDOT \partial_p \, \Gamma_\pi$, its gluon exchange kernel alone can accommodate  the derivative, and \Eq{eq:<x>gB_1loop_int} becomes the  2-loop expression 
\begin{align}
\langle x \rangle_{\rm g_B} =  {\rm tr}  \int_{p,q}^R   \bar{\Gamma}_\pi &
\, S_{\rm u}(p) \, \gamma_\mu \, S_{\rm u}(p\!-\!q)\,  \big(\frac{q \CDOT n}{P \CDOT n}\big) \, \hat{{\mathcal D}}_{\mu \nu}(q)  \nonumber \\ &  \Gamma_\pi\, S_{\rm d}(p\!-\!q\!-\!P)\, \gamma_\nu \,  S_{\rm d}(p\!-\!P) ~,
\label{eq:<x>gB_2loop}
\end{align}
illustrated in the top panel of \Fig{fig:<x>gBind_Pi_m1}.   Here \mbox{$ \hat{{\mathcal D}}_{\mu \nu}(q) =$} \mbox{$ - n \CDOT \partial_{\rm q} \, \hat{ {\mathcal K} }_{\mu \nu}(q) $}, and $q \CDOT n/P \CDOT n$ is the momentum fraction of the exchanged gluon.  
The  binding gluon contribution $\langle x \rangle_{\rm g_B}$ thus completes the sum rule for a purely valence quark pion \mbox{$\langle x \rangle_{\rm q} + \langle x \rangle_{\rm g_D} + \langle x \rangle_{\rm g_B} =$} \mbox{$ 1$}. 

Depending on the details of the non-perturbative DSE-RL model used to generalize the bare relation \mbox{$\hat{{\mathcal D}}^0_{\mu \nu}(q) = $} \mbox{$ {\mathcal K}^0_{\rm g}(q^2)\, \hat{D}_{\mu \nu}(q) $} for the gluon kernel, the two vertices in \Eq{eq:q+g_m1_Vertex} will be model-dependent, but weakly so.  This is because the Ward Identity vertex or propagator is well-constrained through DSE-RL modeling of many quark structure aspects of hadrons.   As a result \mbox{$\langle x \rangle_{\rm Ward} = $}  \mbox{$\langle x \rangle_{\rm q} + \langle x \rangle_{\rm g_D} $} is expected to be very weakly model dependent and just less than unity.   
Within the DSE-RL approach, we obtain for $[\pi, N]$ the $\langle x \rangle_{\rm Ward}$ results  $[0.965, 0.974]$, thus identifying \mbox{$\langle x \rangle_{\rm g_B} = $} \mbox{$[0.035, 0.026]$}.  Since the quark-in-quark  fractions after adjustment for the Wilson line are $[0.602,0.588]$,  the binding gluon accounts for just 5-10 \% of the total gluon momentum fraction, emphasizing the dominance of the gluon dressing effect. 

\begin{table}[tbp]
\ra{0.9}

\begin{tabular}{c|ccc|c||cc}\hline
                   & \hspace{2mm} $J_{\rm u_{\rm v}}$ \hspace{1mm} &  $ J_{\rm d_{\rm v}} $\hspace{1mm} & \hspace{1mm}   $ J_{\rm g}$ \hspace{1mm}   & \hspace{1mm} $J_{\rm tot}$   \hspace{1mm}  & \hspace{2mm} $J_{\rm \bar u}$ \hspace{1mm} & \hspace{1mm} $ J_{\rm \bar d}$  \\
\hline
\rule{0em}{3ex}                                         
    v only       & 0.478  & -0.119       &   0.119      & 0.478    &  0      &     0    \\
 \rule{0em}{3ex} 
                       & 100\%  & -25\%      &   25\%      & 100\%    &  0      &     0     \\
\end{tabular}
\begin{tabular}{c|ccc|c||cc}\hline
                   & \hspace{2mm} $J_{\rm u+\bar u}$ \hspace{1mm} &  $ J_{\rm d+\bar d} $\hspace{1mm}  & \hspace{1mm} $ J_{\rm g}$ \hspace{1mm}   & \hspace{1mm} $J_{\rm tot}$   \hspace{1mm}  & \hspace{2mm} $J_{\rm \bar u}$ \hspace{1mm} & \hspace{1mm} $ J_{\rm \bar d}$ \hspace{1mm}    \\
\hline
\rule{0em}{3ex}  
      $ \!\!{\rm v}\!\!+\!\!\pi N$     & 0.439 & -0.051   &   0.119    &  0.507   &  0.010  & 0.052    \\  
      \rule{0em}{3ex}
                                      & 86.6\% & -10.1\%   &   23.5\%    &  100\%   &  2.1\%  & 10.3\%    \\  
\hline 
\end{tabular}
\caption{ Parton sharing of the proton  $J$ at the model scale $\mu_0 = 0.64$~GeV.   The first section at the top is from the previously established DSE-RL model containing only valence quarks and the dynamically involved dressing glue.  The next section shows the quark $J$ values after pion cloud dressing of the proton involving $\pi^+ n$ and $\pi^0 p$ contributions~\cite{Thomas:2007bc}.
}
\label{tab:JvaluesWilson}
\end{table}

\begin{table}[tbp]
\ra{0.9}

\begin{tabular}{c|ccc|cc|c|cc}\hline 
                   & \hspace{2mm} $L_{\rm u_v}$ \hspace{1mm} &  $ L_{\rm d_v} $\hspace{1mm}   &   $ L_{\rm g} $\hspace{2mm}       &   \hspace{2mm} $ \Delta_{\rm g}$ \hspace{1mm}   & \! $\Sigma_{\rm q}/2$   \!  &  \hspace{1mm}  $J_{\rm tot}$ \hspace{1mm}  &  \hspace{0mm} $L_{\rm \bar u}$ & \hspace{1mm} $L_{\rm \bar d}$ \!\!   \\
\hline
\rule{0em}{3ex}    
      v only      & 0.133    & -0.033    &    0.041       &     0.078          &  0.259      &   0.478      &    0      &     0       \\
\rule{0em}{3ex}    
                     & 27.7\%   & -6.9\%    &    8.6\%      &     16.3\%        &  54\%         & 100\%      &  0         &   0        \\
\end{tabular}
\begin{tabular}{c|ccc|cc|c|cc}\hline 
                   & \hspace{2mm} $L_{\rm u +\bar u}$ \hspace{1mm} &  $ L_{\rm d +\bar d} $\hspace{1mm}   &   $ L_{\rm g} $\hspace{2mm}       &    \hspace{1mm} $ \Delta_{\rm g}$ \hspace{1mm}   & $\Sigma_{\rm q}/2$   &  \hspace{1mm}  $J_{\rm tot}$ \hspace{1mm}   & \hspace{2mm} $L_{\rm \bar u}$ & \hspace{1mm} $L_{\rm \bar d}$ \!\!   \\
\hline
\rule{0em}{3ex}   
          $ \!\!\!{\rm v}\!\!+\!\!\pi N$  & 0.167 & 0.031   &  0.041     & 0.078    &  0.190   & 0.507   &  0.010    &  0.052   \\
\rule{0em}{3ex}    
                                     & 33\% & 6.2\%   &  8.1\%     & 15\%    &  37\%   & 100\%   &  2.1\%    &  10\%   \\ 
\hline 
\end{tabular}
\caption{ Parton sharing of the proton  $L$ at the model scale $\mu_0 = 0.64$~GeV.   Top section is from the previously established DSE-RL model containing valence quarks dressed dynamically via gluons.  Bottom section adds pion cloud dressing of the proton involving $\pi^+ n$ and $\pi^0 p$ contributions.  $L_{\rm g}$ is deduced after separating the canonical helicity \mbox{$S_{\rm g} \approx \Delta_{\rm g} $} as discussed in the text.
}
\label{tab:Lvalues}
\end{table}

\smallskip
\noindent\textbf{Model Nucleon Matrix Element:} 
The nucleon matrix element in \Eq{eq:DSE_spinF_N_q} involves a \mbox{$Q \geq 0$} generalization of the model used previously to study $\langle x \rangle$  sharing between quarks and gluons~\cite{Freese:2021zne}.  Here
\begin{align}
\mathcal{M}^f_{\uparrow, \downarrow}(p,Q_{\perp};\!P)  =  \left[ u_{\downarrow}(p_-\!) \, \bar{u}_{\uparrow}(p_+\!) \right] \rho_{f,\uparrow, \downarrow} \, A(p,Q_{\perp}\!;\!P),
\end{align}
where \mbox{$p_\pm = p \pm Q_{\perp}/2$} and $\rho_{f,\uparrow, \downarrow}$ is the relevant spin-flip quark density for  flavor $f$ in the pure SU(6) spin-isospin proton state.  The diagonal spin probabilities yield the quark numbers  \mbox{$ \small\sum_s \, \rho_{f,s,s} = n_f $}, while  the off-diagonal probabilities are \mbox{$\rho_{u,\uparrow,\downarrow} =\frac{4}{3} $} and \mbox{$\rho_{d,\uparrow,\downarrow} = - \frac{1}{3} $}.  
The amplitude $A$ is
\begin{align}
A(p,Q_{\perp}\!;\!P)= N \,\frac{ \small\sum_{\alpha\prime \alpha} \, f_{\alpha^\prime}(q^\prime)\, f_\alpha(q)}{ ( (K-p)^2 + M_{\rm D}^2) }~,
\label{eq:Nucl_Ampl}
\end{align}
where \mbox{$\alpha = 1, 0 $} label the spin of the spectator pair of quarks, \mbox{$f_\alpha(q) = N_\alpha/( q^2 + R^2_\alpha ) $}, and \mbox{$q^\prime = p_+ \!- \!P^\prime \!/3$} and \mbox{$q = p_- \!- \!P/3$} are the relative momenta of the active quark and  spectator pair in the final and initial state respectively.   The ratio $N_1/N_0 $ replicates the relative infrared strength of the spin-1 and spin-0 $qq$ correlations within the Faddeev amplitudes employed  in \Ref{Bednar:2018htv}, while $N$ is determined by valence quark number.
\begin{figure}[tbp]
\centering\includegraphics[width=\columnwidth,height=58mm]{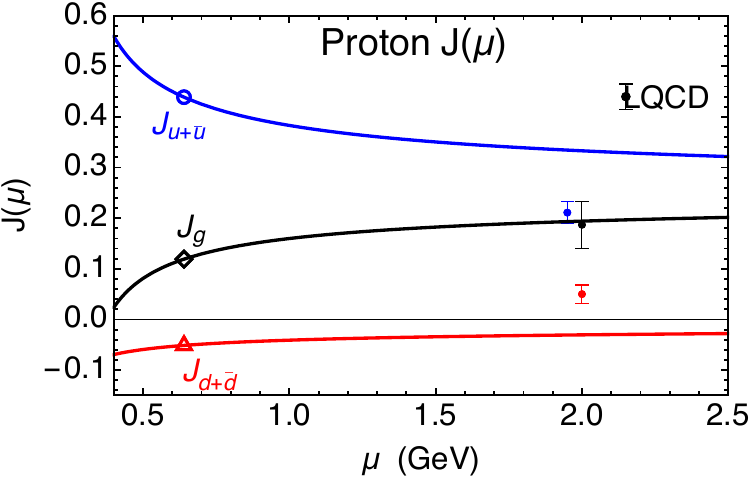}
\caption{Approximate evolution with scale of the parton shares of the proton $J$.  Symbols show the values at model scale \mbox{$\mu_0 = 0.64~{\rm GeV} $} that include sea quark contributions from the pion cloud mechanism.   The nature of the evolution approximation is discussed in the text.  The displayed LQCD results at  \mbox{$\mu=2$}~GeV are from \Ref{Alexandrou:2017oeh}.  
}
\label{fig:Nucl_J_scaling_wWilson}  
\end{figure}

\smallskip 
\noindent\textbf{Rainbow-Ladder Model Results}:
The DSE-RL gluon kernel ${\mathcal K}_{\rm g}(q^2)$ generates the gluon momentum fraction and its strength was previously set~\cite{Freese:2021zne} so that $\langle x \rangle_{\rm g}^{\pi}(\mu_0)$ (binding plus dressing effects lumped together), together with the Landau gauge valence and sea $\langle x \rangle_{\rm q}^{\pi}(\mu_0)$, led to a minimized RMS departure from global data analysis results at upper scales.  Here we wish to employ the earlier estimated 7\% reduction to both valence and sea $\langle x \rangle_{\rm q}^{\pi}(\mu_0)$ to account for the
Wilson line and this requires a compensating 27\% increase in $\langle x \rangle_{\rm g}^{\pi}(\mu_0)$ to maintain the fit to data at higher scales.~\footnote{If one used the -17\% quark reduction suggested by the recent 1-loop analysis with strictly bare coupling~\cite{Costa:2021mpk} then the required $\langle x \rangle_{\rm g}^{\pi}(\mu_0)$ would be significantly larger than the global analysis result at 2~GeV.   This bare coupling estimate is not compatible with the infrared running couplings and dynamical propagators of the DSE-RL approach.  }
The new gluon kernel  strength parameter $ D_{\rm gg}$, along with that of the unchanged quark interaction kernel, are summarized in an Appendix.   Because these adjustments to  $\langle x \rangle_{\rm q}(\mu_0)$ and  $\langle x \rangle_{\rm g}(\mu_0)$ are of opposite sign and very nearly equal in magnitude,  a good fit to upper scale global data analysis after NLO DGLAP evolution is achieved by an increase in the identified model scale $\mu_0$.  After the pion fixes the parameters, they are applied unchanged to the nucleon.  We note that the fitted $\langle x \rangle_{\rm sea}(\mu_0)$ values include non-valence flavors as needed.  Results shown in \Table{Pion-N_m1_mu0_mu} also include $\langle x \rangle_{\rm q,g}$ at  selected higher scales compared to data analysis results. 
 
The $\pi-N$ cloud mechanism~\cite{Thomas:2007bc}  is an overall 2\% contribution to \mbox{$\langle x \rangle(\mu_0)$}~\cite{Bednar:2018htv}; however it is a 6\% contribution to $J(\mu_0)$, so we display those results here.   To implement this, the valence-only and the P-wave $\pi-N$  Fock sectors for the spin-up proton $p$ are given weights $(1-Z;Z)$ with  the latter divided into $(\pi^+,n;\pi^0,p)$ sectors according to the isospin coupling weights.   Then each isospin sector is separated by angular momentum coupling weights into sectors having the core nucleon $J$ projection up or down. 
The Fock space weight $Z=0.201$ is determined, as in \Ref{Bednar:2018htv}, by the Gottfried Sum Rule for the flavor asymmetry of sea quark numbers. 

In \Table{tab:JvaluesWilson} we display the $J$ values obtained for the u- and d-quarks and gluon at the model scale.  The top section employs only valence quarks, while the bottom section includes the pion cloud mechanism~\cite{Thomas:2007bc} which generates $J_{\bar q}$ and modifies $J_{q}$.   The valence only  $J_{\rm tot}$ is $4.4\%$ too low; with a realistic inclusion of the pion cloud it improves to be just $1.4\%$ too high.   It is to be expected that a model will not produce a perfect result for total $J$; here Lorentz covariance is maintained but only a limited infinite class of Feynman diagrams has been employed, and the pion cloud model is grafted on to the basic valence model.  It is not a perfect fit; but the final difference of just $1.4\%$ suggests that the dominant physics has been accounted for.  It is worth noting that the $u$ quark and gluon totally dominate over the $d$ quark, providing 87\% and 24\% of $J_{\rm tot}$ respectively; in particular \mbox{$J_{\rm g} \approx 2\, |J_{\rm d + \bar d}| $}.  This minor role of the $d$ quark increases with scale, as seen below.   

In \Table{tab:Lvalues} we display the separate $L_{\rm q}$ and \mbox{$S_{\rm q} =\Sigma_{\rm q}/2$} values  at the model scale, both with and without the pion cloud mechanism.   Since a gauge-invariant expression for $S_{\rm g}$ within the Ji or kinetic approach via GPDs has not been established~\cite{Ji:1996ek,Ji:1996nm,Leader:2013jra}, we approximate it via the canonical helicity, i.e.,  \mbox{$S_{\rm g} \approx \Delta_{\rm g} $} to then provide the qualitative estimate  \mbox{$L_{\rm g} \approx $}  \mbox{$J_{\rm g} - \Delta_{\rm g}$} as suggested in \Ref{Ji:1996ek}.  The quark and gluon helicities are gauge invariant and have been obtained from the second Mellin moments of the polarized PDFs as discussed previously within this present DSE-RL approach and nucleon model~\cite{Freese:2021zne}.   \Table{tab:Lvalues} also displays the resulting $L_{\rm g}$.  The indications are that $J_{\rm tot}$ is shared almost equally between the total intrinsic spin \mbox{$\Delta_{\rm g} +\Sigma_{\rm q}/2$} and the estimated total orbital contribution.   The d-quark orbital contribution is minor, about 6\%.  The obtained 37\% contribution of the quark intrinsic spin agrees with the modern consensus, while the 15\% contribution from gluon helicity at model scale is not easily ignored.  The significant gluon contributions to $J, L, S$ suggested  here extend previous model investigations of nucleon $J$ that did not include gluon dynamics at model scale, see e.g., \Refs{Thomas:2007bc,Freese:2020mcx}. 
\begin{figure}[h]
\centering\includegraphics[width=\columnwidth,height=58mm]{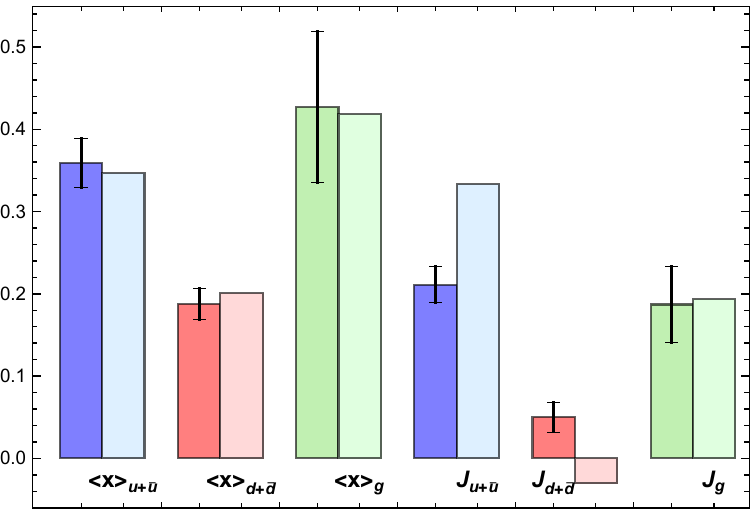}
\caption{ Values for parton momentum fractions and angular momenta  at  \mbox{$\mu = 2~{\rm GeV} $} compared to  lattice QCD results at physical quark masses~\cite{Alexandrou:2020sml}. 
}
\label{fig:BarChart_N_x_J_2GeV_2020}  
\end{figure}

\begin{figure}[h]
\centering\includegraphics[width=\columnwidth,height=58mm]{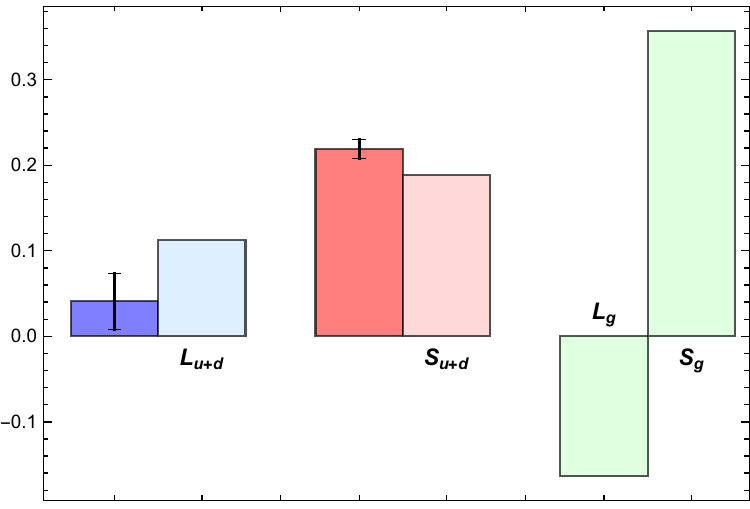}
\caption{$L$ and $S$ decomposition for valence quarks and glue at  \mbox{$\mu = 2~{\rm GeV} $} compared to  lattice QCD results for quarks at physical quark masses as reported by \Ref{Alexandrou:2020sml}.  As explained in the text, the model results for the glue decomposition required an extra approximation. 
}
\label{fig:BarChart_N_u_d_2GeV_2020}  
\end{figure}

The LO scale evolution of the parton components of $J$ is carried out with the expressions provided in \Refs{Ji:1995cu,Thomas:2008ga} and the results at \mbox{$\mu = 2$}~GeV are displayed in 
\Fig{fig:Nucl_J_scaling_wWilson}  compared with  recent results from lattice QCD~\cite{Alexandrou:2017oeh}.  The very small $J_{\rm s+\bar s}$ component generated from zero at the model scale is not shown.
A caution is needed.  The scale evolution expressions of \Refs{Ji:1995cu,Thomas:2008ga} are for the canonical or Jaffe-Manohar~\cite{Jaffe:1989jz}  components of $J$ which are not gauge-invariant~\cite{Leader:2013jra}. 
In \Table{tab:JL_2GeV_2020}, \Fig{fig:BarChart_N_x_J_2GeV_2020},  and \Fig{fig:BarChart_N_u_d_2GeV_2020} we provide the results at 2~GeV including the separate $L_{\rm q}$ and $S_{\rm q}$  components compared to a recent LQCD work.~\footnote{ The total $J$ of our model at $2$~GeV is slightly lower than the value at $\mu_0$ in \Table{tab:JvaluesWilson} because the evolution equations of \Ref{Ji:1995cu} preserve scale invariance of total $J$ only if the starting scale result is perfect ($1/2$). }      Similar to analysis of the model scale,  we again adopt the helicity approximation  \mbox{$S_{\rm g} \approx \Delta_{\rm g} $} to then provide the qualitative model estimate for the gluon  \mbox{$L_{\rm g} \approx $}  \mbox{$J_{\rm g} - \Delta_{\rm g}$} as suggested in \Ref{Ji:1996ek}.  The parton momentum fractions at 2~GeV are also included in \Fig{fig:BarChart_N_x_J_2GeV_2020} and the similar agreement is not surprizing considering they are all second Mellin moments. 
The quark results are heavily constrained by the closely related first Mellin moments associated with conservation of baryon and flavor number.  The gluon results are heavily influenced by the gluon dressing of quarks that implements dynamical chiral symmetry breaking.  The quark-in-quark and gluon-in-quark mechanisms are tied together dynamically.
Within the present DSE-RL approach we have produced the total $\langle x \rangle_{\rm g}$ and $J_{\rm g}$ from the dressing mechanism and have not attempted to separate out  "binding gluon" contributions.   As discussed  earlier, a robust estimate for $\langle x \rangle_{\rm g_B} $ in the pion and nucleon  is at most 10\% of $\langle x \rangle_{\rm g_{tot}}$.   A similar fraction is expected for the "binding gluon" contribution to $J_{\rm g}$.

Our estimate for $L_{\rm g}$  is strong, almost $-1.5 L_{\rm  q}$, and is due to the significant increase of $\Delta_{\rm g}$ with scale, from $0.078$ at the model scale to $0.357$ at $\mu = 2$~GeV.   However the  $L_{\rm g}$ values here should be treated with caution considering the assumption used to separate it from $J_{\rm g}$. 
\begin{table}[tbp]
\ra{0.9}

\begin{tabular}{c|cc|cc|cc}\hline 
                   & \hspace{2mm} $J$ \hspace{1mm} &  $J_{\rm LQCD}$ \hspace{2mm} & \hspace{2mm} $L$ & \hspace{0.2mm} $L_{\rm LQCD}$  \hspace{0.5mm}  & \hspace{1mm}   $ S $ \hspace{1mm}       &    \hspace{-0.3mm} $  S_{\rm LQCD} $     \\
\hline
\rule{0em}{3ex}   
                   ${\rm u+\bar u}$        & 0.333 (66\%)    &   0.211 (44\%)   &  0.061   &  -0.221     &     0.272         &    0.432               \\
\rule{0em}{3ex}    
                    ${\rm d+\bar d}$       &  -0.030 (-6\%)   &  0.050 (11\%)    &  0.052    &  0.262       &   -0.083       &    -0.213              \\ 
\rule{0em}{3ex}    
                    ${\rm h+\bar h}$       &  0.008  (2\%)    &  0.025  (5\%)    &  0.008   &  0.053       &     0               &   -0.028              \\ 
\rule{0em}{3ex}    
                    ${\rm g}$                   &  0.194  (38\%)    &  0.187 (40\%)    &  -0.163   &      -            &    0.357         &        -              \\ 
 \hline
\rule{0em}{3ex}    
                  {\rm Tot}                    &   0.504              &  0.473                &  -0.042    &      -            &    0.546          &       -                \\                     
\hline 
\end{tabular}
\caption{ Present results (including the $\pi N $ cloud mechanism) for parton sharing of the proton  $J$, $L$ and intrinsic spin $S = \Sigma_{\rm q}/2 +\Delta_{\rm g}$ at  $\mu = 2$~GeV compared to LQCD results~\cite{Alexandrou:2020sml}.   That LQCD work did not separate $J_{\rm g}$ into $L_{\rm g}$ and $S_{\rm g}$.  
}
\label{tab:JL_2GeV_2020}
\end{table}
The enhanced gluon role at this higher scale is evidenced by  \mbox{$J_{\rm g} \approx 6\, |J_{\rm d + \bar d}| $} and \mbox{$\Delta_{\rm g} \approx 71\% \, J_{\rm tot}$} within the present approach.  Since the canonical scale evolution~\cite{Altarelli:1977zs,Ji:1995cu} gives the asymptotic results \mbox{$J_{\rm g}(\mu \to \infty) = 16/50 $}, \mbox{$\Delta_{\rm g} (\mu \to \infty) = \infty $}, and thus \mbox{$L_{\rm g} (\mu \to \infty) = -\infty $}, the sensible measure of the gluon contribution remains $J_{\rm g}$. This is 24\% of $J$ at model scale and 38\% at $2$~GeV, indicating the importance of gluon dynamics for accurate angular momentum considerations at any scale.   
 
Our results agree with earlier studies in that $L_{\rm u}$ dominates over $L_{\rm d}$ in models at low scale and this tends towards a reversal as the scale increases.   By 2~GeV a reversal is clear in  LQCD results~\cite{Alexandrou:2020sml,Alexandrou:2017oeh,Hagler:2004er} and in some models~\cite{Thomas:2008ga,Freese:2020mcx}.   Here we find  equality \mbox{$L_{\rm u} \!= \!L_{\rm d}$} is reached only at \mbox{$\mu \approx 2.5$}~GeV. 
The isovector  $L_{\rm u} \!- \!L_{\rm d}$ should be reliable in lattice calculations due to the absence of disconnected contributions.    
Our result for the isoscalar $L_{\rm u} \!+ \!L_{\rm d}$ is $0.11$ which compares well to both \Ref{Freese:2020mcx} and the LQCD  value $0.041$ shown in \Table{tab:JL_2GeV_2020}.  

\smallskip 
\noindent\textbf{Summary}:
We have explored the dynamically coupled quark and gluon contributions to both $J$ and $\langle x\rangle$ for the proton, as obtained from the second Mellin moments of the unpolarized GPDs.   Our emphasis has been on exploring the mechanisms underlying the distribution of $J$ among partons that can be revealed in a first application of the Rainbow-Ladder DSE model.  The strong u- and d-quark dressing underlying dynamical mass generation provides quark-in-quark and gluon-in-quark mechanisms that dominate the observed quark and gluon proportions of $J$ and $\langle x\rangle$ for the proton.  Insight gained from a strictly 1-loop treatment of a quark target, 
have provided estimates of the Wilson line contribution to both $J$ and $\langle x\rangle$ within the Landau gauge DSE-RL approach.  We also identify the "binding gluon" contribution $\langle x \rangle_{\rm g_B}$ as a 2-loop integral for the pion.  Since the GPDs directly access the $J$ contributions from each parton, including gluons, this accounting for nucleon spin improves on historical modeling approaches wherein quark helicities are first obtained from SU(6) wavefunctions, separate models are used to generate $L_{\rm q}$ information, and $J_{\rm g}$ is ignored or input as phenomenology.  

The GPDs of the DSE-RL  approach give a good account of nucleon $J$.   The gluon component is found to be   \mbox{$J_{\rm g} \sim 24\%$} at $\mu_0$, and $\sim 38\%$ at 2~GeV.  It is more important than the d-quark contribution, being  \mbox{$J_{\rm g}(\mu_0) \approx 2\, |J_{\rm d + \bar d}| $} at $\mu_0$ and $6$ at $2$~GeV.   
The Wilson line correction to the Landau gauge $J_{\rm q}$ was estimated to be $-7\%$ using insight from a 1-loop model;  in the DSE-RL approach it is found to be effectively incorporated by an increase in the assigned model scale.    
The "binding gluon" contribution to  $\langle x \rangle_{\rm g}$ for the pion was found to be $\leq 10\%$.  This is consistent with the dynamical chiral symmetry breaking quark dressing being so strong that the gluon-in-quark mechanism provides most of $\langle x \rangle_{\rm g}$ and $J_{\rm g}$.   Comparison with LQCD at 2~GeV would be more meaningful with availability of evolution equations for $J_{\rm q/g}(\mu)$ that match the Ji or kinetic definitions.


\smallskip 
\noindent\textbf{Appendix: DSE Model Parameters}:
Parameters of the Landau gauge interaction kernels of \Eq{eq:RL_Kernel} and \Eq{eq:CutG_Kernel} and the model nucleon amplitude of \Eq{eq:Nucl_Ampl} are summarized here.  (To respond to  the estimated Wilson line effect, $ D{\rm gg}$ has increased as explained in the text.)
\begin{table}[h]
\ra{0.9}

\begin{tabular}{c|cc|cc}\hline
              &\hspace{2mm} $D_{\rm RL} $ \hspace{1mm} &  $\omega$\hspace{1mm} & \hspace{1mm} $ D_{\rm gg}$  &  \hspace{1mm} $ \omega_{\rm g}$         \\
\hline
 ${\mathcal K}(q^2)\,, {\mathcal K}_{\rm g}(q^2)$ \hspace{1mm} &  37.324  & 0.5 &  3.18 & 0.53     \\
 \hline  
             &\hspace{2mm}  $ M_{\rm D} $ \hspace{1mm}    & \hspace{1mm} $ N_1/N_0 $ \hspace{2mm} & \hspace{1mm} $ R_{\rm s}$   \hspace{1mm} &  $ R_{\rm v}$         \\      
 \hline    
  $A_N(p,P)$  &    1.005  &       -1.28      &     0.5       &      0.8               \\
\hline
\end{tabular}
\caption{Model parameters. }
\label{Params}
\end{table}

\smallskip 
\noindent\textbf{Appendix:  Kernel for $J_{\rm g}$}:
The gluon kernel  $\hat{{\mathcal D}}_{\mu \nu}(q,Q_{\perp})$ appearing in \Eq{eq:Inhom_GVertm} is 
${\mathcal K}_{\rm g}(q) \, \hat{D}_{\mu \nu}(q,Q_{\perp})$ with the second factor being the tensor
\begin{align}
\hat{D}_{\mu \nu} =  & 2\, q \CDOT n \Big(\!\delta_{\mu \nu} \!- \!\frac{ n(q_+\!)_\mu \, q_{+}(q_-\!)_\nu + q_{-}(q_+\!)_\mu \, n(q_-\!)_\nu }{q \CDOT n}  \nonumber \\
& + \frac{ n(q_+\!)_\mu \, (q_+ \!\CDOT q_-\!) \, n(q_-\!)_\nu }{ ( \!q \CDOT n \!)^2}  \nonumber \\
& + \frac{ q_{-}(q_+\!)_\mu \, q_{+}(q_-\!)_\nu - {q_{-}}_\mu \, {q_{+}}_\nu }{ (\!q_+ \CDOT q_-\!) }  \!\Big)~,
\label{eq:Inhom_XgQ}
\end{align}
where \mbox{$q_\pm = q \pm Q_{\perp}/2$}, and the projection of $a$ perpendicular to $v$ is denoted  \mbox{$a(v)_\mu = $} \mbox{$a_\mu - v_\mu (v \CDOT a) / v^2 $}.  
Note that \mbox{$\hat{D}_{\mu \nu}(q,Q_{\perp} \!\to \!0) = \hat{D}_{\mu \nu}(q)$} given in \Eq{eq:D_explicit}.

\smallskip 
\noindent\textbf{Appendix: Form of quark vertices for $J_{\rm q/g}$:} 
The quark vertex $\Gamma_{\rm q}^{(m)}(p,Q_\perp\!) $ defined in \Eq{eq:Gamma_BSE}, and that provides the contribution to $J_{\rm q}$ when $m=1$, has the covariant form 
\begin{align}
\Gamma_{\rm q}^{(m)}& =  (\frac{p \CDOT n}{P \CDOT n} )^m\,  i \big[ \sigma(n Q_\perp\!)  F_1^{(m)}  \!+  \sigma(n p) \, p \CDOT Q_\perp  \!F_2^{(m)}  \nonumber \\
+  & \sigma(p  \,Q_\perp\!) \, p \CDOT n\,  F_3^{(m)}  \nonumber \\ 
+  &( \nslash \,\pslash  \,\Qperpslash \!+ n \CDOT Q_\perp \pslash - \!p \CDOT n  \,\Qperpslash \!- p \CDOT Q_\perp \nslash  )\,  F_4^{(m)}  \big],
\label{eq:q_J_Vert}
\end{align}
where only the necessary contributions linear in $Q_\perp$ have been retained and the amplitudes $F_i^{(m)}(p^2)$ are real functions of $p^2$.   In the above we have used the notation \mbox{$\sigma(a b) =\sigma^{\mu \nu} a_\mu b_\nu$}.   The quark vertex $\Gamma_{\rm g}^{(m)}(p,Q_\perp\!)$ that carries the dressing gluon contribution to  $J_{\rm g}$ has the same form.

\smallskip
\noindent\textit{Acknowledgments:}
We are grateful to A. W. Thomas for beneficial discussions on nucleon spin.   


%

\end{document}